# LATTICE COMPUTATION OF A MAGNETIC MONOPOLE MASS[*]

J. SMIT

*Institute for Theoretical Physics, University of Amsterdam*
*Valckenierstraat 65, 1018 XE Amsterdam, The Netherlands*

and

A.J. VAN DER SIJS

*Department of Theoretical Physics, University of Zaragoza*
*Facultad de Ciencias, 50009 Zaragoza, Spain*

ABSTRACT

A single magnetic monopole in pure SU(2) gauge theory is simulated on the lattice and its mass is computed in the full quantum theory. The results are relevant for our proposed realization of the dual superconductor hypothesis of confinement.

## 1. Introduction

In Ref. 1 we presented a realization of the dual superconductor hypothesis of confinement in SU(2) gauge theory. The monopoles were of the 't Hooft-Polyakov (HP) form, in particular the Bogomol'nyi-Prasad-Sommerfield (BPS) solution. Assuming that the configurations considered dominate the string tension $\sigma$ we obtained the estimate $\sqrt{\sigma}/\Lambda_{\overline{MS}} = 2.3$.

In our model we made two assumptions. The first one was that quantum fluctuations screen the (chromo-)electric field of the classical BPS monopole, turning it into what we call an HP-like monopole. The second assumption concerned the mass of the monopole. The mass of the classical BPS monopole of scale $\mu$ is $M = 4\pi\mu/g^2$. We conjectured that the monopole mass in the full quantum theory behaves as

$$M = \frac{4\pi\mu}{g_R^2(\Lambda_R/\mu)} C(g_R^2(\Lambda_R/\mu)) . \qquad (1)$$

Here $g_R$ is the running coupling in the $R$-scheme defined in terms of the quark-antiquark potential, with $\Lambda_R/\Lambda_{\overline{MS}} = 1.048$. The function $C$ was supposed to be slowly increasing as for the HP monopole, $1 < C \lesssim 2$.

In the present study, described in detail in Ref. 2, we investigate the second assumption numerically. We put a single monopole in a box, by imposing appropriate monopole boundary conditions, and use lattice Monte Carlo methods to include quantum fluctuations. The simulation data are used to determine the function $C(g_R^2(\Lambda_R/\mu))$ which allows us to check the validity of the mass formula (1).

## 2. The magnetic monopole

The classical magnetic monopole of (arbitrary) scale $\mu$ in euclidean pure SU(2) gauge theory is given by the gauge field configuration

$$A_k^a(\vec{x},t) = \epsilon_{akl}\hat{x}_l \frac{1-K(\mu r)}{r}, \qquad A_4^a(\vec{x},t) = \delta_{ak}\hat{x}_k \frac{H(\mu r)}{r}, \qquad (2)$$

with

$$H(\mu r) = \mu r \frac{\cosh \mu r}{\sinh \mu r} - 1, \qquad K(\mu r) = \frac{\mu r}{\sinh \mu r}, \qquad (3)$$

as in the BPS limit of the HP monopole.

The idea is to induce the monopole in a finite cubic spatial box (times a periodic time direction of length $T$) by fixing the fields in the boundary at a value suggested by the asymptotic behaviour of the monopole field. We take 'HP-like' boundary conditions with parameter $\mu_0$ for our dynamical simulations, i.e. $K \sim 0$, $H/r \sim \mu_0$. An analysis of classical monopole energies shows that this is compatible with both an HP monopole of scale $\mu = \mu_0$ and a BPS monopole of scale $\mu_{\text{eff}} = \mu_0 + 1/R_{\text{eff}}$. Another important result of the classical analysis is that the accessible range of $\mu_0$ values is restricted to $0 < \mu_0 \leq \pi/T$. This is due to a symmetry implying that $\mu_0$ and $\mu_0'$ boundary conditions are equivalent if $\mu_0' = \mu_0 + 2\pi n/T$, for some integer $n$, and to the occurrence of monopoles of opposite electric charge.

## 3. The monopole mass

The monopole mass can be written as the sum of a contribution from inside the box,

$$M_{\text{in}} = -\frac{1}{T} \ln \frac{Z_{\text{mon}}(\beta; a\mu_0)}{Z_{\text{vac}}(\beta)}, \qquad (4)$$

measured in the simulation, and a correction term for the outside region. Here

$$Z_{\text{mon}}(\beta; a\mu_0) = \int DU \exp[-S_\Box(U;\beta)] = \int DU \exp[-\beta \sum_{\text{plaquettes}} (1 - \frac{1}{2} \text{tr } \Box(U))] \qquad (5)$$

is the partition function subject to monopole boundary conditions and $Z_{\text{vac}}(\beta)$ is the analogous definition for vacuum boundary conditions, i.e. $A_\mu^a = 0$. The gauge coupling constant $g$ is embodied in $\beta = 4/g^2$ and $\beta \to \infty$ is the classical limit.

Eq. (4) can be written in a form accessible to Monte Carlo computation by differentiating it with respect to $\beta$ and subsequently integrating again,

$$M_{\text{in}}(\beta) = \int_0^\beta \frac{d\tilde{\beta}}{\tilde{\beta}} \Delta E(\tilde{\beta}) = \int_0^\beta \frac{d\tilde{\beta}}{\tilde{\beta}} \frac{1}{T}(\langle S \rangle_{\text{mon}} - \langle S \rangle_{\text{vac}}). \qquad (6)$$

In order to compute this numerically the integral is replaced by a sum. At each value of $\tilde{\beta}$ in this summation two simulations are needed, to compute $\langle S \rangle$ with both monopole and vacuum boundary conditions. High statistics is required to compute the difference of these two large numbers accurately.

Subsequently, $C_{\text{in}}(g_R^2(\Lambda_R/\mu))$ is extracted from the $M_{\text{in}}$ data as follows. First $\Lambda_R/\mu$ is calculated using Monte Carlo data for $a\sqrt{\sigma}$ and taking a value for $\sqrt{\sigma}/\Lambda_R$. Next $g_R^2(\Lambda_R/\mu)$ is calculated using its two-loop $\beta$-function. Inserting this into Eq. (1) leads to $C_{\text{in}}(g_R^2(\Lambda_R/\mu))$. This procedure is carried out for both $\mu = \mu_0$ and $\mu = \mu_{\text{eff}}$, corresponding to HP-like and BPS-like behaviour of the monopole, respectively.

Fig. 1 shows $C$ for a simulation at an $8^4$ lattice, with $\mu_0 = \pi/8 = 0.39$. The correction due to the mass contribution from the outside region is included. For $g_R^2 \approx 0$ we know the monopole is BPS-like, so the HP analysis (upper three curves) is misleading there. At larger couplings, the monopole may or may not become HP-like. If it does, the $C$ values will lie in the region indicated by the upper set of curves. This means that $C$ increases from $C = 1$ at weak coupling to $C \approx 1.6$ at $g_R^2 \approx 8$ or $C \approx 2.0$ at $g_R^2 \approx 10$, depending on the value of $\sqrt{\sigma}/\Lambda_R$. This is in good agreement with our assumptions. If, however, the monopole remains BPS-like at large coupling, our first assumption does not apply. Nevertheless, even in that scenario (lower set of curves) $C$ remains almost constant, $C \approx 1$ up to $g_R^2 \approx 6$.

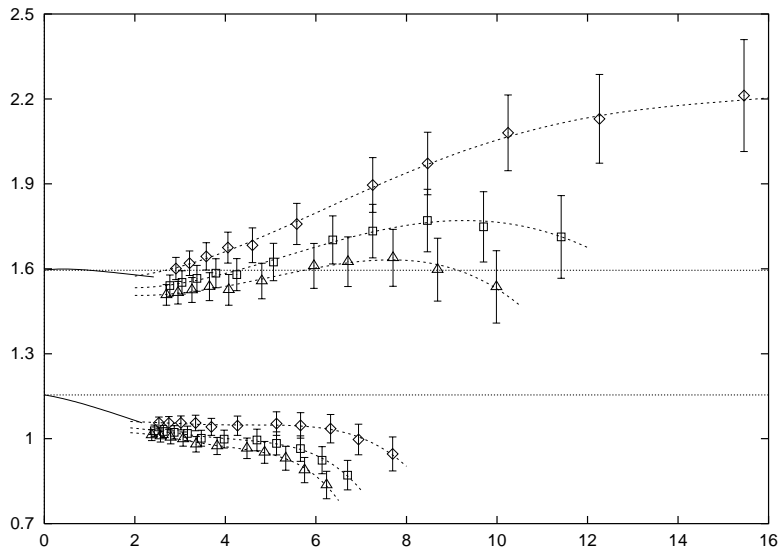

Fig. 1. $C$ as a function of $g_R^2(\Lambda_R/\mu)$. The different sets of points are for $\sqrt{\sigma}/\Lambda_R = 1.7$ ($\Diamond$), 2.0 ($\Box$) and 2.2 ($\triangle$), for both $\mu = \mu_0$ (upper set of curves) and $\mu = \mu_{\text{eff}}$ (lower set). The horizontal lines denote the classical limit ($g^2 = 0$). The solid curves come from large-$\beta$ fits of the mass data.

**Acknowledgements:** This work was carried out when AvdS was at the University of Oxford, supported by SERC (U.K.) grant GR/H01243. He is currently sponsored by DGICYT (Spain). JS is supported by the Stichting FOM. The numerical simulations were performed on the Cray Y-MP4/464 at SARA with support from the Stichting NCF.